\documentclass[10pt,twocolumn]{article}

\usepackage[T1]{fontenc}
\usepackage[utf8]{inputenc}
\usepackage[margin=0.75in]{geometry}
\usepackage{mathptmx} 

\usepackage{graphicx}
\usepackage{amsmath}
\usepackage{amssymb}
\usepackage{booktabs}
\usepackage{multirow}
\usepackage{array}
\usepackage[normalem]{ulem}
\usepackage[shortlabels]{enumitem}
\usepackage{listings}
\usepackage{xcolor}
\usepackage{color}
\usepackage[numbers,sort&compress]{natbib}
\usepackage{hyperref}

\lstdefinestyle{bashstyle}{
    language=bash,
    basicstyle=\ttfamily\small,
    breaklines=true,
    frame=single,
    columns=fullflexible,
    showstringspaces=false,
    keepspaces=true,
    backgroundcolor=\color{gray!5},
    rulecolor=\color{gray!30}
}
\begin{document}

\twocolumn[
\begin{center}
  {\LARGE\bfseries BIP! Ranker: A Software Library for Citation-Based Impact Indicators on Large-Scale Graphs\par}
  \vspace{1.2em}
  {\large
    Ilias Kanellos$^{1,\dagger}$ \quad
    Serafeim Chatzopoulos$^{2,\ast}$ \quad
    Thanasis Vergoulis$^{2}$
  \par}
  \vspace{0.6em}
  {\normalsize
    $^{1}$Independent Researcher, Athens, Greece \quad
    $^{2}$IMSI, Athena Research Center, Athens, Greece
  \par}
  \vspace{0.4em}
  {\small
    $^{\dagger}$Work conducted while affiliated with IMSI, ATHENA RC. \quad
    $^{\ast}$Corresponding author: \texttt{schatz@athenarc.gr}
  \par}
\end{center}
\vspace{1.2em}

\noindent\textbf{Abstract.} Scientific impact is multidimensional: overall influence, current popularity, early citation momentum, and field-relative performance each capture a distinct facet of a publication's impact. Yet, in practice, these dimensions are often reduced to a single metric, such as citation count. Open solutions for computing multiple complementary impact indicators at scale remain scarce, particularly for citation graphs as large as those provided by major scholarly databases. We introduce BIP! Ranker, an open-source, Spark-based library for computing citation-based impact indicators at scale, capable of processing citation networks with billions of citations among hundreds of millions of publications.

\vspace{0.8em}
\vspace{1.5em}
]

\section{Introduction}
\label{sec:intro}

Citation analysis plays a central role in scientific knowledge discovery and scientometrics. Modern scholarly discovery services often combine keyword relevance with citation-based indicators to rank search results and help users navigate the vast volume of research indexed in their databases. Citation-based indicators are also valuable for a wide range of scientometric analyses involving scientific impact estimation, including the identification of emerging fields or the study of scientific collaboration and knowledge diffusion.

In most cases, a simple indicator, such as citation count, is used as a proxy for scientific impact. However, scientific impact is inherently multidimensional. Total citation count reflects only a publication's cumulative influence from its publication date to the present, and cannot capture other important dimensions in many contexts, such as current popularity, early citation momentum, or impact relative to the norms of its research field.
For example, biology and mathematics have markedly different citation distributions, making raw citation counts difficult to compare across disciplines. Likewise, when identifying recent research trends, total citation count is an inadequate indicator: citations received many years ago are not reliable proxies for a work's current attention, and recently published papers cannot be meaningfully compared with older publications that have had much more time to accumulate citations. Over the past decades, several indicators have been introduced to capture these distinct dimensions of scientific impact. Examples include RAM~\cite{ghosh2011tar}, ECM~\cite{ghosh2011tar}, and AttRank~\cite{kanellos2021attrank} for popularity and FWCI~\cite{purkayastha2019comparison} for accounting for disciplinary differences in citation practices.

However, despite evidence of their usefulness in a variety of experiments~\cite{kanellos2019survey}, many of these indicators have not been widely adopted by scientific knowledge-discovery systems and remain uncommon in scientometric analyses. Although some software libraries support the calculation of simple metrics, such as total citation count, reliable open-source implementations of more advanced indicators remain scarce.
Even for citation counts, most systems and analyses rely on precomputed values provided by well-established scholarly datasets. These values are frequently generated using restricted or proprietary software, with limited disclosure of the data-processing choices and assumptions that determine them. This lack of transparency can compromise the reproducibility and interpretability of downstream systems and analyses.
Turning to advanced indicators, even the relatively few available open-source implementations often cannot scale to the largest citation networks, containing billions of citations across hundreds of millions of publications, like those available through open scholarly datasets such as the OpenAIRE Graph~\cite{manghi_paolo_2022_7488618} and OpenAlex~\cite{priem2022openalex}.

In this paper, we present \emph{BIP! Ranker}, an open-source software library for the scientometrics and bibliometrics community that enables the efficient computation of publication-level, citation-based impact scores on large-scale citation networks. It implements a diverse set of indicators capturing influence, popularity, citation momentum, and field-normalised impact, enabling scholarly search and discovery services to present multiple complementary views of research impact.

\section{Related Work}
\label{sec:related}

The literature has introduced a range of impact indicators capturing different aspects of scientific impact, such as AttRank~\cite{kanellos2021attrank} and RAM/ECM~\cite{ghosh2011tar}. The papers introducing such indicators do not always release open implementations, and when they do, these are typically not designed to scale to large citation graphs. Moreover, they exist as separate, independently maintained implementations, so building a system that computes multiple indicators means hand-picking and integrating implementations from different sources. BIP! Ranker instead brings multiple indicator families together in a single, open-source library that scales to large citation networks, removing the need for this ad hoc integration.

General-purpose graph processing tools implement fundamental algorithms such as PageRank but are not designed specifically for scholarly impact computation. Single-node libraries, including NetworkX~\cite{hagberg2020networkx} and igraph~\cite{csardi2013package}, are convenient for prototyping, but their in-memory execution limits applicability to citation networks with millions of nodes and billions of citations. Distributed frameworks such as Apache Spark's GraphX~\cite{gonzalez2014graphx} address this scalability limitation but offer only general-purpose graph primitives, leaving domain-specific implementations to application developers. BIP! Ranker builds on distributed graph computation while adding the layers required for scalable scholarly impact analysis: a shared citation-graph representation, tested computation parameters and conventions, and execution mechanisms for computing multiple validated impact indicators.

On the other hand, interactive citation analysis tools such as VOSviewer~\cite{vaneck2010vosviewer}, CitNetExplorer~\cite{vaneck2014citnetexplorer}, and the \textit{bibliometrix} R package~\cite{aria2017bibliometrix} target exploratory analysis and visualisation rather than large-scale computation, typically assuming data that fits on a single machine. They also lack a production-oriented interface for computing and integrating multiple indicators.

\section{Library Description}
\label{sec:design}

\subsection{Main Components and Interface}

\begin{figure}[t]
    \centering
    \includegraphics[width=\linewidth]{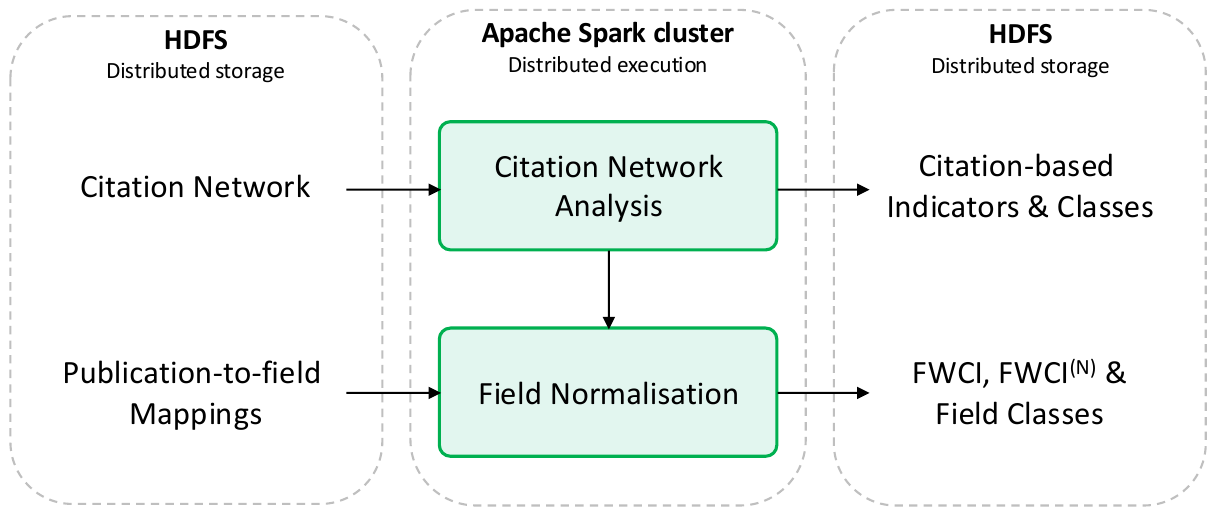}
    \caption{BIP! Ranker's inputs, main components, and outputs.}
    \label{fig:arch}
\end{figure}

Figure~\ref{fig:arch} illustrates BIP! Ranker's main components, along with their inputs and outputs. The \emph{Citation Network Analysis} component operates directly on the citation network, computing all citation-based impact indicators described in Section~\ref{sec:graph-indicators}. It is implemented as a series of Apache Spark (PySpark) scripts that can run on a computational cluster to analyse very large citation networks, producing publication-level scores together with percentile-based impact classes. Both input and output files are stored on a distributed HDFS storage.

The \emph{Field Normalisation} component derives field-normalised indicators by combining the citation-based scores produced by the Citation Network Analysis component with a provided publication-to-field mapping. It computes Field-Weighted Citation Impact (FWCI)~\cite{purkayastha2019comparison} and field-specific impact classes for all indicators produced by the Citation Network Analysis component. The component is also implemented in Spark and its inputs and outputs are stored on HDFS.

\subsection{Implemented Indicators}
\label{sec:graph-indicators}

BIP! Ranker implements citation-based impact indicators spanning the major indicator families summarised in Table~\ref{tab:indicators}. Indicators in the \emph{Influence} family capture a publication's overall, long-term scientific impact, based on its complete citation history. Indicators in the \emph{Popularity} family reflect the attention a publication receives today, thereby addressing the bias against recently published articles caused by citation lag~\cite{diodato2013dictionary}. Indicators in the \emph{Impulse} family capture the momentum an article gains in its early years, distinguishing publications that attract attention soon after publication from ``hidden gems'' that gain recognition with delay. Finally, \emph{Field-normalised} indicators adjust citation scores relative to the relevant research field, thereby reducing the effect of differences in citation practices across research areas.
The following paragraphs provide further details on each implemented citation-based impact indicator.

\begin{table}[b]
  \caption{Citation-based impact indicators implemented in BIP! Ranker.}
  \label{tab:indicators}
  \footnotesize
  \begin{tabular}{@{}lll@{}}
    \toprule
    \textbf{Family} & \textbf{Indicators} & \textbf{Implementation Script} \\
    \midrule
    Influence & CC, PageRank & \texttt{CC.py}, \texttt{PageRank.py} \\
    Popularity & RAM, ECM, AttRank & \texttt{TAR.py}, \texttt{AttRank.py} \\
    Impulse & iCC & \texttt{CC.py} \\
    Field-normalised & FWCI, $N$-year FWCI & \texttt{TopicClassesAndFWCI.py} \\
    \bottomrule
  \end{tabular}
\end{table}

\paragraph{Citation Count (CC)}
The most widely used impact indicator, CC aggregates the citations a publication has received over its entire citation history. Formally, the CC score of publication $i$ is given by
\begin{equation}
\label{eq:cc}
CC_i = \sum_j A_{i,j}
\end{equation}
where $A$ is the adjacency matrix of the citation graph, i.e., $A_{i,j}=1$ if publication $j$ cites publication $i$, and $A_{i,j}=0$ otherwise. It offers a simple, cumulative proxy for overall influence, though it treats every citation as equally important regardless of the impact of the citing works.

\paragraph{PageRank}
BIP! Ranker implements the standard random-surfer PageRank formulation~\cite{page1999pagerank}, exposing the damping factor and convergence threshold as configurable parameters, with default values chosen to perform well on citation networks. The score of publication $i$ is computed as
\begin{equation}
\label{eq:pagerank}
PR_i = \alpha \sum_j P_{i,j}\, PR_j + (1-\alpha)\,\frac{1}{|V|}
\end{equation}
where $P$ is the stochastic transition matrix (the column-normalised version of $A$), $\alpha \in [0,1]$ is the damping factor, and $|V|$ is the number of publications (vertices) in the citation graph. The indicator propagates scores across the citation graph, assigning greater influence to citations originating from highly cited publications. As an influence indicator, it complements the previously discussed CC by considering the network's entire structure rather than only direct citation in-links. This helps identify foundational publications that may not be highly cited themselves but are cited by influential, highly cited publications.
To support large-scale computation of this indicator, the implementation checkpoints intermediate Spark DataFrames during iterations, preventing excessive lineage growth and enabling processing of large citation graphs.

\paragraph{RAM and ECM}
RAM and ECM implement the time-aware citation model of \citet{ghosh2011tar}, which discounts citations by the recency of the citing paper. RAM applies this weighting directly to individual incoming citations, effectively extending CC with a recency-based decay factor:
\begin{equation}
\label{eq:ram}
RAM_i = \sum_j R_{i,j}, \qquad
R_{i,j} =
\begin{cases}
\gamma^{\,t_c - t_j} & \text{if } j \text{ cites } i \\
0 & \text{otherwise,}
\end{cases}
\end{equation}
where $R$ is the Retained Adjacency Matrix, $\gamma \in (0,1)$ is a decay parameter, $t_c$ is the current year, and $t_j$ is the publication year of citing article $j$.
ECM generalises this approach to entire citation chains, weighing each chain by both its length and the age of the citations composing it:
\begin{equation}
\label{eq:ecm}
ECM_i = \sum_j \sum_{h=1}^{H} \lambda^{h} \left(R^{h}\right)_{i,j}
\end{equation}
where $R^h$ denotes the $h$-th power of the retained adjacency matrix $R$, $\lambda \in (0,1]$ is a chain-length attenuation parameter that discounts longer citation chains, and $H$ is the maximum chain length considered. RAM represents the single-hop case ($H=1$) of ECM's chain-based propagation, allowing both indicators to share a common computation pipeline.

\paragraph{AttRank}
AttRank~\citep{kanellos2021attrank} extends PageRank by modifying its teleportation component to incorporate attention signals derived from publication and citation recency. The AttRank score of publication $i$ is computed as
\begin{equation}
\label{eq:attrank}
AttRank_i = \alpha \sum_j P_{i,j}\, AttRank_j + \beta \cdot Att(i) + \gamma \cdot c \cdot e^{-\rho\,(t_c - t_i)}
\end{equation}
where $\alpha + \beta + \gamma = 1$ with $\alpha,\beta,\gamma \in [0,1]$, $Att(i)$ is a recent attention-based score reflecting publication $i$'s share of citations in recent years,
$t_i$ is its publication year, $t_c$ is the current year, $c$ is a normalisation constant, and $P$ is the stochastic transition matrix.
AttRank captures short-term scientific attention while preserving the graph-based score propagation mechanism of PageRank. It is classified as a popularity measure, complementing the time-aware citation indicators RAM and ECM.

\paragraph{``Incubation'' Citation Count (iCC)}
iCC is a time-restricted variant of CC that counts only the citations a publication receives within a fixed window of $N$ years after its publication (typically a small value, e.g., $N=3$), rather than over its entire citation history:
\begin{equation}
\label{eq:icc}
iCC_i^{(N)} = \sum_{j:\, t_j \le t_i + N} A_{i,j}
\end{equation}
where $A$ is the adjacency matrix and $t_i$, $t_j$ are the publication years of the cited and citing papers, respectively. It thus captures a publication's early citation momentum (``impulse''), during an initial citation window.
Since iCC restricts the citation window considered by CC, it follows the same underlying computation approach while applying an additional temporal constraint.

All the aforementioned indicators are implemented as Apache Spark (PySpark) scripts that produce publication-level scores, together with categorical, percentile-based classes: publications are ranked in descending order according to their indicator scores and assigned to one of five classes---Top 0.01\%, Top 0.1\%, Top 1\%, Top 10\%, or Average (the remaining 90\%)---reflecting their relative standing within the underlying publication collection. Such classes are commonly used by scientific discovery services to offer users a more intuitive way to assess a publication's standing.

Beyond the global percentile-based classes, BIP! Ranker also provides field-level impact classes. Publications are first assigned to their relevant research fields using mappings provided as input. The percentile-assignment procedure is then applied independently within each field-specific publication set.
This allows publications to be assessed relative to other papers in the same field,
reducing the effect of differences in citation practices across research
areas.

Moreover, the citation-based indicators (CC and iCC) can be combined with
publication-to-field mappings in a second processing stage to compute
field-normalised indicators. Specifically, Field-Weighted Citation Impact
(FWCI) is computed following its standard definition~\cite{purkayastha2019comparison}:
\begin{equation}
\label{eq:fwci}
\mathrm{FWCI}_i =
\frac{\mathrm{CC}_i}
{\mathbb{E}\!\left[\mathrm{CC}\mid d_i,\, y_i,\, f_i\right]}
\end{equation}
where $d_i$, $y_i$, and $f_i$ denote the document type, publication year, and
assigned research field of publication $i$, respectively. The denominator
therefore represents the expected citation count for publications sharing the
same document type, publication year, and research field.

In addition, BIP! Ranker provides an $N$-year FWCI variant that applies the
same normalisation procedure while restricting both observed and expected
citations to those received within the first $N$ years after publication:
\begin{equation}
\label{eq:nfwci}
\mathrm{FWCI}_i^{(N)} =
\frac{iCC_i^{(N)}}
{\mathbb{E}\!\left[iCC^{(N)} \mid d_i,\, y_i,\, f_i\right]}
\end{equation}
where $iCC_i^{(N)}$ denotes the $N$-year incubation citation count of
publication $i$, as defined in Eq.~\ref{eq:icc}.

\subsection{Large-Scale Distributed Processing}
\label{sec:scale}

BIP! Ranker is designed to compute citation-based indicators over publication collections that exceed the capacity of a single machine. By exploiting Apache Spark's distributed
processing model, it can process citation networks containing hundreds of millions of publications and billions of citation relationships, including networks of the scale encountered in large scholarly knowledge graphs.

To demonstrate this capability, we executed the indicator-compu-tation pipeline on a citation network extracted from the OpenAIRE Graph~\cite{manghi_paolo_2022_7488618} comprising $385$M publications and $2.1$B citation relationships. The experiments were performed on an Apache Spark cluster managed by Hadoop YARN, with 32 worker nodes providing a total of 2.44~TB of memory and 1,280 vCPUs.

The framework completed both simple aggregation-based indicators and iterative graph-based algorithms at this scale. Citation Count (CC) was computed in approximately $24$ minutes, while PageRank and AttRank completed in approximately $79$ and $72$ minutes, respectively. These results illustrate that BIP! Ranker can leverage distributed computing resources to support the practical computation of citation-based indicators over large-scale scholarly networks.

\section{Usage Example}
\label{sec:usage}

BIP! Ranker can be executed locally or in a distributed Spark environment. Detailed installation instructions, software requirements, and command-line arguments are available in the project's README file.\footnote{ \url{https://github.com/athenarc/Bip-Ranker/blob/main/README.md}}
It also includes a demonstration pipeline that executes the complete workflow on a sample citation graph:

\begin{lstlisting}
python run_demo.py
\end{lstlisting}

Each indicator script operates directly on the citation-graph representation described in Section~\ref{sec:design}. For example, Citation Count and PageRank can be computed as follows:

\begin{lstlisting}
python CC.py <graph_path> <num_partitions>

python PageRank.py <graph_path> <alpha> \
  <convergence_error> <checkpoint_dir> \
  <num_partitions> <mode>
\end{lstlisting}

The resulting indicator scores can subsequently be combined with publication-to-field mappings to compute topic-level impact classes and field-normalised indicators:

\begin{lstlisting}
python TopicClassesAndFWCI.py \
  --scores-file <scores.tsv> \
  --concepts-file <pub-to-fields.tsv> \
  --output-dir <output_dir>
\end{lstlisting}

The complete documentation describes all available scripts, configuration options, input/output formats, and deployment scenarios.

\section{Availability}
\label{sec:availability}

BIP! Ranker is publicly available on GitHub\footnote{\url{https://github.com/athenarc/Bip-Ranker}} under the GNU General Public License v2 (GPL-2.0). The software version discussed in this paper corresponds to GitHub release \texttt{v2.0.0}\footnote{\url{https://github.com/athenarc/Bip-Ranker/releases/tag/v2.0.0}} and is permanently archived on Zenodo~\cite{kanellos_2026_20448503}. The repository includes the source code, sample data, documentation, and a minimal example workflow to reproduce the software described in this paper. The source code is maintained under version control using Git.

\section{Impact and Reuse}
\label{sec:impact}

\textbf{Target users.} BIP! Ranker is intended for researchers and practitioners who develop scientific knowledge-discovery services or conduct scientometric analyses on large-scale publication collections. By providing a unified implementation of multiple citation-based indicators, it eliminates the need to reimplement them.

\textbf{Reproducibility.} Each implemented indicator directly follows the corresponding definition presented in Section~\ref{sec:graph-indicators}. Together with the open-source implementation, this enables researchers to examine, validate, and reproduce analyses conducted using the library. Rather than relying on ready-made citation metrics from scholarly databases, whose calculation procedures may not be transparent, researchers can use BIP! Ranker to compute well-documented and reproducible indicators for their own publication collections.

\textbf{Operational use.} BIP! Ranker constitutes the computational backend of several scholarly services developed by our group, including BIP! Finder~\cite{vergoulis2019bipfinder}, BIP4COVID19~\cite{vergoulis2022bip4covid19}, BIP! DB~\cite{vergoulis2021bipdb}, and BIP! Scholar~\cite{vergoulis2022bipscholar}.
These systems collectively demonstrate the applicability of the library across literature search, researcher assessment, and domain-specific discovery services.
In particular, BIP! DB publishes BIP! Ranker's indicator scores computed based on the OpenAIRE Graph~\cite{manghi_paolo_2022_7488618} as an openly available dataset on Zenodo.\footnote{BIP! DB: \url{https://doi.org/10.5281/zenodo.4386934}}

\textbf{Adoption.}
BIP! Ranker's outputs
are also integrated into independent scholarly infrastructures.
The OpenAIRE Graph~\cite{manghi_paolo_2022_7488618} incorporates the computed impact scores and percentile classes into its production workflow, exposing them through the Graph dataset,\footnote{OpenAIRE Graph dataset: \url{https://doi.org/10.5281/zenodo.3516917}}
APIs, and OpenAIRE Explore.\footnote{OpenAIRE Explore: \url{https://explore.openaire.eu/}}
These indicators support ranking functionalities
and provide impact information alongside research products.
Furthermore, the ELIXIR-supported bio.tools registry~\cite{ison2019biotools} displays impact indicators, computed by BIP! Ranker, on software records, demonstrating reuse of the generated metrics beyond the original BIP! ecosystem.

\textbf{Maintenance and sustainability.} BIP! Ranker is maintained by the Scientific Knowledge and Data Management (SKnow) Lab at Athena Research Center.
The public source code repository and open-source licensing facilitate access, reproducibility, and future extensions by the research community.

\section{Conclusion}
\label{sec:conclusion}

We presented BIP! Ranker, an open-source library for computing complementary citation-based impact indicators over large-scale citation graphs. BIP! Ranker contributes a common computational framework for the scalable, consistent integration of established bibliometric measures spanning influence, popularity, impulse, and field-normalised impact. Its value is demonstrated through the sustained use in BIP! Services and the adoption by independent infrastructures, including OpenAIRE and bio.tools. BIP! Ranker thus provides a reusable, field-aware foundation for scholarly ranking, discovery, and research assessment.

\bibliographystyle{unsrtnat}
\bibliography{thebib}

\end{document}